\documentclass{ws-mpla}
\usepackage[super]{cite}
\usepackage{graphicx}
\begin{document}

\markboth{Warren, Mathews, Meixner, Hidaka, \& Kajino}
{STERILE NEUTRINO DARK MATTER IN CORE-COLLAPSE SUPERNOVAE}

\catchline{}{}{}{}{}

\title{IMPACT OF STERILE NEUTRINO DARK MATTER ON CORE-COLLAPSE SUPERNOVAE}

\author{\footnotesize MACKENZIE L. WARREN}

\address{Department of Physics \& Astronomy, Michigan State University,\\
 East Lansing, Michigan 48824, USA\\
\&\\
Department of Physics, Center for Astrophysics \& Joint Institute for Nuclear Astrophysics, University of Notre Dame \\
 Notre Dame, Indiana 46556, USA \\
mwarren@msu.edu}

\author{GRANT J. MATHEWS}

\address{Department of Physics, Center for Astrophysics \& Joint Institute for Nuclear Astrophysics, University of Notre Dame \\
 Notre Dame, Indiana 46556, USA 
}

\author{MATTHEW MEIXNER}
 
 \address{ Space Exploration Sector, The Johns Hopkins University Applied Physics Laboratory,  \\
Laurel, Maryland 20723 USA}

\author{JUN HIDAKA}

\address{Mechanical Engineering Department, Meisei University,\\
2-1-1 Hodokubo, Tokyo 191-8506, Japan}

\author{TOSHITAKA KAJINO}

\address{National Astronomical Observatory of Japan, 2-21-1 Osawa, \\
Mitaka, Tokyo 181-8599, Japan \\
\& \\
Department of Astronomy, Graduate School of Science, The University of Tokyo, 7-3-1 Hongo, \\
Bunkyo-ku, Tokyo, 113-0033, Japan}

\maketitle

\pub{Received (\today)}{Revised (\today)}

\begin{abstract}
We summarize the impact of sterile neutrino dark matter on core-collapse supernova explosions.  We explore various oscillations between electron neutrinos or mixed $\mu-\tau$ neutrinos and right-handed sterile neutrinos  that may occur  within a core-collapse supernova.  In particular, we consider sterile neutrino masses and mixing angles that are consistent with sterile neutrino dark matter candidates as indicated by recent X-ray flux measurements.  We find that the interpretation of the observed 3.5 keV X-ray excess as due to a decaying 7 keV sterile neutrino that comprises 100\% of the dark matter would have almost no observable effect on supernova  explosions.  However, in the more realistic case in which the decaying sterile neutrino comprises only a small fraction of the total dark matter density due to the presence of other sterile neutrino flavors, WIMPs, etc., a larger mixing angle is allowed.  In this case a 7~keV sterile neutrino could have a significant impact on core-collapse supernovae.  We also consider mixing between $\mu-\tau$ neutrinos and sterile neutrinos.  We find, however, that this mixing does not significantly alter the explosion and has no observable effect on the neutrino luminosities at early times.

\end{abstract}

\ccode{PACS Nos.: 95.35.+d, 14.60.St, 15.50.+x, 97.60.Bw,14.60.Pq}
\section{Introduction}

The nature of dark matter remains one of the biggest open questions in physics and is vital to developing a complete picture of both cosmology and fundamental physics.  One viable dark matter candidate is a sterile neutrino \cite{abazajian2001b}.  Sterile neutrino dark matter is an electroweak singlet that  does not participate in the strong, weak, or electromagnetic  interaction.  Thus, it is consistent with limits from the Large Electron-Positron collaboration measurement \cite{LEP2006} of the width of the $Z^{0}$ gauge boson.  The standard model does not provide any predictions about this proposed particle, but  bounds can be placed \cite{abazajian2001a,abazajian2001b,boyarsky2006b,boyarsky2007,boyarsky2009b,boyarsky2009c,boyarsky2014,bulbul2014,chan2014} on the mass and mixing angle parameter space for sterile neutrino dark matter by X-ray astronomy and cosmology, i.e $1\text{ keV} < m_{s}< 18 \text{ keV}$ and $\sin^{2} 2 \theta_{s} \lesssim1.93 \times 10^{-5} \left(\frac{m_{s}}{\text{keV}}\right)^{-5.04}$.

At the same time neutrinos play an intriguing and important role in both cosmology and the dynamics of core-collapse supernovae (CCSNe).  Neutrinos dominate the energetics of core-collapse supernovae due to the high neutrino fluxes that are emitted as the proto-neutron star forms.  Observations indicate that the typical explosion  kinetic energy of CCSNe is $\sim10^{51}$~ergs.   This  is only $\sim$1\% of the emitted neutrino energy within the first few seconds after bounce.  

Despite recent developments in computational methods, however, a detailed understanding  the explosion mechanism of CCSNe has not yet emerged.  Spherically symmetric models tend to not explode and those two-dimensional (2D) and three dimensional (3D) models that do explode, tend to explode with too little energy \cite{janka2012}.  The problem is that, although a shock forms successfully and propagates outward in mass, it loses energy to the photodissociation of heavy nuclei and quickly becomes a standing accretion shock.  

There have been many attempts to revitalize the stalled shock via hydrodynamic effects, such as the standing accretion shock instability (SASI)\cite{blondin2003,blondin2006,scheck2008} and neutrino heated convection \cite{herant1994,burrows1995,janka1996,murphy2013}, but even in these cases the explosion energies tend to be too low \cite{janka2012}.  Nevertheless, given the enormous neutrino fluxes present, it is unavoidable that neutrinos should play a dominant role in the explosion mechanism.

Even one dimensional spherically symmetric models will explode in simulations with enhanced neutrino fluxes, either from convection below the neutrinosphere \cite{wilson1988,wilson1993,book} or a QCD phase transition \cite{fischer2011}.  In this paper, we explore the resonant mixing \cite{hidaka2006,hidaka2007} between a sterile and electron  (or antineutrinos) as a means  to increase the early neutrino luminosity and revitalize the shock.  In Ref.~[\refcite{warren2014}] it was found that  enhanced explosion energies could occur for a broad range of  masses and mixing angles.  Even spherical models that do not explode could be made to explode by the presence of mixing to sterile neutrinos.  The present work, however, will focus specifically on masses and mixing angles that are consistent with constraints on sterile neutrino dark matter candidates.   We also explore other mixing possibilities than just $\nu_{e} \leftrightarrow \nu_{s}$ oscillations.

The efficient transport of neutrino energy using sterile neutrinos was first proposed by Hidaka and Fuller \cite{hidaka2006,hidaka2007}.  Using a one zone collapse calculation, they studied the resonant oscillations of a sterile neutrino with the mass and mixing angle of a warm dark matter candidate.  A resonant conversion $\nu_{e}\rightarrow \nu_{s}$ could occur deep inside the proto-neutron star core.  Although this initially decreases the shock energy by decreasing the electron fraction in the core and removing energy, they found that a second resonance may occur near the  pro-neutron star surface.  Thus, the resonant mixing of sterile neutrinos and electron neutrinos can serve as an efficient method of transporting neutrino energy out from the proto-neutron star core, where high energy neutrinos are trapped.  The sterile neutrinos then arrive at the  the stalled shock to assist in neutrino reheating.  This mechanism is highly sensitive to the feedback between neutrino oscillations and the local composition, energy transport, and hydrodynamics and warrants  detailed numerical studies such as those summarized in this paper.

In Ref.~[\refcite{warren2014}] coherent active-sterile neutrino oscillations were studied using the University of Notre Dame-Lawrence Livermore National Laboratory (UND/LLNL) code \cite{book,bowers1982}, a spherically symmetric  relativistic  hydrodynamic supernova model with flux limited multi-group  neutrino transport. Sterile neutrinos with masses and mixing angles consistent with dark matter constraints and  their impact on shock reheating were studied.  It was found that sterile neutrino dark matter candidates can enhance the shock energy and lead to a  successful explosion.

In Ref.~[\refcite{warren2014}] a large region of the sterile neutrino parameter space  was considered that varied over several orders of magnitude in mass and mixing angle.  It was concluded  that much of the parameter space led to efficient neutrino reheating and resulted in successful explosions.  In this  paper we   look more specifically at the parameter space allowed for dark matter candidates and also consider $\nu_{s}$ mixing with $\nu_{\mu,\tau}$.

{\bf Sterile neutrinos in core-collapse supernovae have also been considered in the context of proto-neutron star cooling and heavy element nucleosynthesis. Limits can be placed on the sterile neutrino mass and mixing angle based upon energy loss arguments in CCSNe \cite{nunokawa1997,arguelles2016}.  These works invoke a single MSW resonance structure between active and sterile neutrino flavors in order to constrain the mixing parameters using observations of SN1987A.  Additionally, much work has been done on the impact of sterile neutrinos in heavy element nucleosynthesis in the neutrino driven winds \cite{nunokawa1997,mclaughlin1999,caldwell2000,fetter2003,tamborra2012,pllumbi2015}.  The research on nucleosynthesis has focused on a light $\sim$eV mass sterile neutrino that is consistent with anomalous results found at the LSND and MiniBooNe reactor neutrino experiments and relies upon a single resonance to convert $\nu_{e} \rightarrow \nu_{s}$ in the neutrino driven wind in order to generate the low electron fraction necessary in the region for r-process nucleosynthesis to occur.  The present work, however, will focus on the impact of $\sim$keV mass sterile neutrinos.  These experience a unique double resonance effect  in the  proto-neutron star core and do not result in a net energy loss to a sterile neutrino species during PNS cooling.}

The organization of this manuscript is as follows.  In Sec.~\ref{sec:dm} we discuss the theory and possible observational signatures of sterile neutrino dark matter.  In Sec.~\ref{sec:osc} we discuss the active-sterile matter-enhanced resonances. In Sec.~\ref{sec:results}, we show that the straightforward interpretation of the observationally inferred 7 keV sterile neutrino accounting for 100\% of dark matter would have a mixing angle too small to affect CCSNe.  However, in a more realist scenario in which the total DM is comprised of a significant contributions from other neutrino flavors, WIMPS, etc, an enhanced CCSNe explosion is possible.  Some discussion and conclusions are provided in Sec.~\ref{sec:conc}.

\section{Sterile neutrino dark matter \label{sec:dm}}

It seems unnatural that in the standard model there are no right handed neutrinos.  This is after all the simplest symmetry imaginable.  If neutrinos are  massless, and lepton numbers are conserved as was thought when the standard model was first developed, then one must avoid a right-handed neutrino.  However, it is now known that neutrinos both have mass and that  they  change their flavors. Hence, the introduction of right handed (sterile neutrinos) is, perhaps the most plausible extension of the standard model. One simply adds to the standard model Lagrangian a Dirac equation for the right-handed neutrinos $N_I$ plus Yukawa couplings $F_{\alpha I}$,\cite{boyarsky2014}
\begin{equation}
{\cal L} = {\cal L}_{SM}  + i\bar N_I\partial_\mu \gamma^\mu N_I - \biggl( F_{\alpha I} \bar L_\alpha N_I \tilde \phi - \frac{M_I}{2} \bar N_I^c N_I + b.c.\biggr) ~~,
\end{equation} 
where $M_I$ is the sterile neutrino mass, $L_{\alpha}$ are left-handed leptonic doublets (where $\alpha = e,\, \mu,\, \tau$), $\phi$ is a Higgs doublet, and $F_{\alpha I}$ are Yukawa couplings.  Since the right handed neutrinos have no electric, weak or strong charges they are dubbed ``sterile'' or ``singlet''.

Introducing this Lagrangian has the possibility to solve two of the problems of the standard model, i.e. the masses of active neutrinos and the existence of dark matter.
If the Dirac masses $M_D = F_I \langle \phi \rangle$  are much smaller than the Majorana masses $M_I$, then  the type I seesaw formula \cite{minkowski1977, ramond1979, mohapatra1980,yanagida1980,schechter1980,schechter1982} relates the active neutrino masses $m_\nu$ to the sterile neutrino masses, i.e.
\begin{equation}
(m_\nu )_{\alpha \beta} = - \sum_{I = 1}^N (M_D)_{\alpha I} \frac{1}{M_I} (M_D^T)_{I \beta} ~~,
\end{equation}
where $(m_\nu)_{\alpha \beta}$  is a $3 \times  3$ matrix of active neutrino masses, mixings, and (possible) CP-violating phases.

Also, since there is freedom to select from a wide range of  sterile neutrino masses, they have the possibility to be  a viable dark matter candidate.
However, detecting such sterile neutrinos in dark matter detectors is challenging compared to that of other WIMP candidates since its interaction with normal matter is 
suppressed by a factor of the weak coupling constant times the oscillation mixing angle with normal matter.  Rather, the best chance to detect sterile neutrino dark matter is through its 
radiative decay or its affect on core collapse supernovae which is the subject of this  paper.

Radiative decays of dark matter are allowed in many models, including sterile neutrinos.   In extensions of the standard model, sterile neutrinos have several possible decay modes.  The primary decay channel is to three light, active neutrinos and is thus virtually invisible.  However, several radiative decay channels are also allowed.  The most important is via two body decay to a photon and a light, active neutrino, e.g.~$\nu_{s} \rightarrow \nu_{e} + \gamma$, where  the photon energy is simply $E_{\gamma} = m_{s}/2$.   The rate of this decay has been investigated for both Dirac and Majorana neutrinos \cite{pal1982} and is large enough to potentially place bounds on the sterile neutrino mass and mixing angle.  The rate for a Majorana neutrino is \cite{pal1982}
\begin{equation}
\Gamma_{\gamma}(m_{s},\theta_{s}) = \frac{ 9 \alpha G_{F}^{2}}{256 \cdot 4\pi^{4}} \sin^{2} 2 \theta \,m _{s}^{5} \approx 1.38\times10^{-29} \text{s}^{-1} \left(\frac{\sin^{2} 2 \theta_{s}}{10^{-7}}\right)\left(\frac{m_{s}}{1 \text{keV}}\right)^{5},
\label{eq:gamma}
\end{equation}
where $G_{F}$ is the Fermi coupling constant, $\sin^{2} 2\theta_{s}$ is the singlet neutrino mixing angle and $m_{s}$ is the sterile neutrino mass.  It should be noted, however, that this decay rate is model dependent \cite{pal1982} and only applies for the most basic Majorana neutrino mass models.

With the advent of modern X-ray telescopes, the detection of such decay modes is feasible.  Although the decay rate is small, the dark matter density in galactic halos  and galaxy clusters is large enough to generate a potentially measurable flux.  Assuming that sterile neutrinos are the only component of dark matter, the flux observed in a detector from a dark matter halo should scale as \cite{boyarsky2006b}
\begin{equation}
F_{DM}=\frac{M_{DM}}{8 \pi D^{2}} \frac{ \Gamma_{\gamma}}{m_{s}}\sim \frac{M_{DM} \sin^{2} 2 \theta_{s} m_{s}^{4}}{8\pi D^{2}},
\label{eq:flux}
\end{equation}
where $D$ is the distance to the source, $\Gamma_{\gamma}(m_{s},\theta_{s})$ is the decay rate for a sterile neutrino of mass $m_{s}$ and mixing angle $\theta_{s}$ (Eq.~\ref{eq:gamma})), and $M_{DM}$ is the total dark matter mass within the telescope's field of view. 
If the observed decaying sterile neutrinos are only a fractional component of the total dark matter  (due to other sterile neutrino flavors, WIMPS, etc), then a particular observed flux may be fit by a larger mixing angle, since the flux given in Eq.~(\ref{eq:flux}) scales with the product of both the total observed mass of decaying sterile neutrino and sterile neutrino mass and mixing angle.  As the fraction of the total dark matter mass in decaying sterile neutrino diminishes, a larger mixing angle is required to explain the observed flux. 

\begin{figure}[t!]
   \centerline{\includegraphics[width=\textwidth]{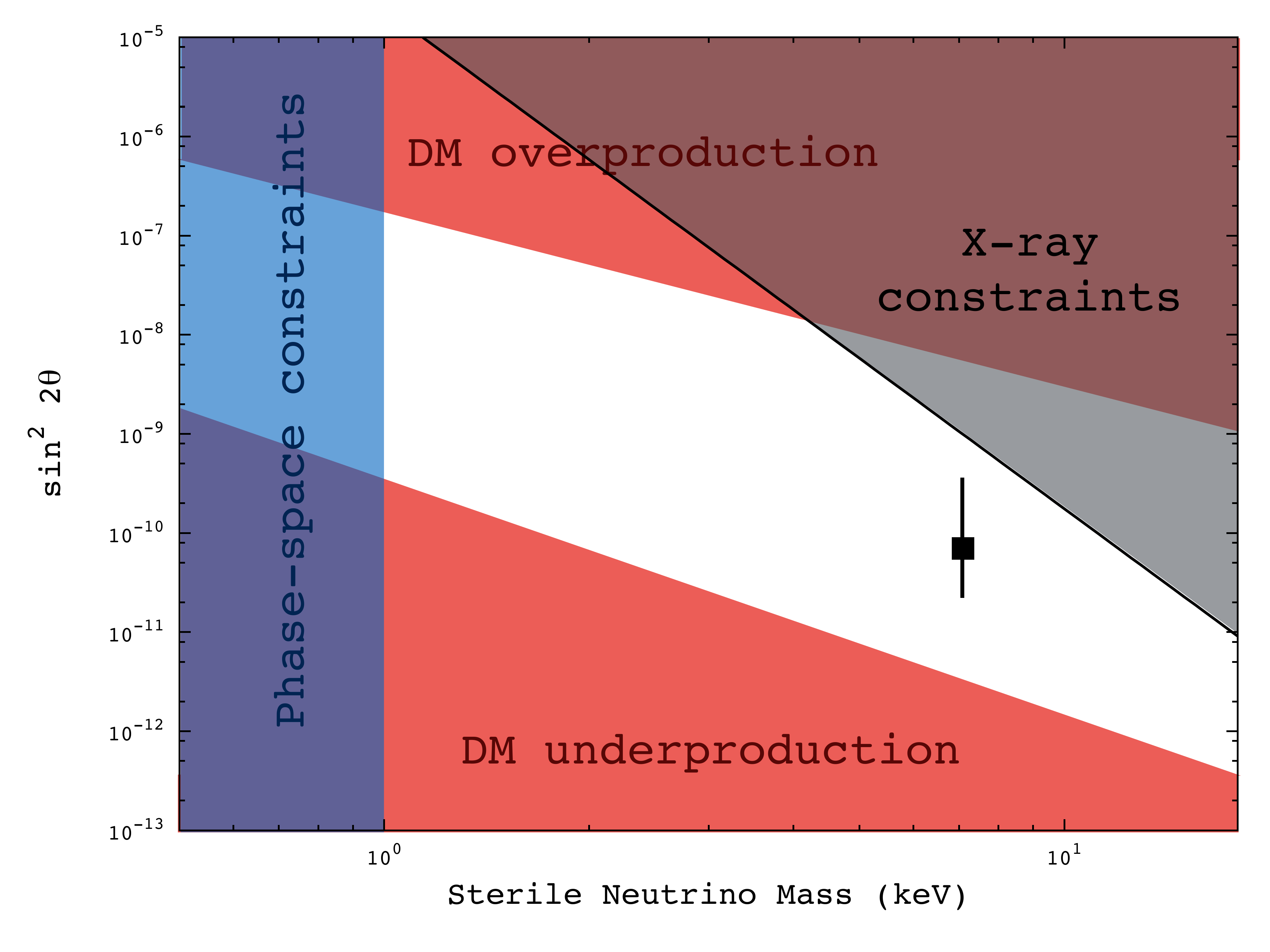}}
   \vspace*{8pt}
   \caption{Bounds on sterile neutrino dark matter.  Phase-space constraints and Lyman-$\alpha$ constraints \cite{tremaine1979,boyarsky2009a,boyarsky2009c} are shown in blue and dark matter over- and under-production \cite{boyarsky2014} in red.  The grey shaded region shows the current exclusion limits from stacked X-ray observations \cite{abazajian2001a,boyarsky2006a,boyarsky2006b,boyarsky2007,boyarsky2009c} and the solid black square shows the most recent best fit point from the X-ray flux\cite{boyarsky2014,bulbul2014}.  (Color available online)}
   \label{fig:bounds}
\end{figure}

Observations with the \textit{Chandra} and \textit{XMM Newton} satellites of X-ray emission from various galaxies and galaxy clusters \cite{abazajian2001a,boyarsky2006a,boyarsky2006b,boyarsky2007,boyarsky2009c} have attempted to put upper bounds on the parameters of these decays.   However, recent observations by Boyarsky et al.\cite{boyarsky2014} and Bulbul et al.\cite{bulbul2014} have detected an unidentified emission line at $\sim$~3.5 keV.  The mixing angle is best fit as $\sin^{2} 2 \theta \approx 7 \times 10^{-11}$  for a single dark-matter sterile neutrino.   Additional bounds may be placed on the mixing of a sterile neutrino with an electron neutrino from dark matter over- and under-production in the $\nu$MSM \cite{boyarsky2014} and phase space bounds for fermonic dark matter \cite{tremaine1979,boyarsky2009a,boyarsky2009c}.  These bounds are summarized in Figure \ref{fig:bounds}, alongside the current limits from X-ray observations.  

Interpretation of this anomalous  X-ray emission \cite{bulbul2014,boyarsky2014} however remains ambiguous due to possible observational backgrounds.  Many dark matter systems, such as the galactic center or galaxy clusters, are X-ray bright objects with contributions from both point sources and plasma emission.  Although efforts have been made to model this background using multi-temperature plasma models, constraining uncertainties such as the temperatures and compositions in such complicated systems is difficult.  It has been argued \cite{bulbul2014,jeltema2015} that the observed 3.5 keV line may be due to known plasma lines, mainly K XVIII lines at 3.48 and 3.52 keV.  However, fits of the 3.5 keV emission with a potassium line  \cite{jeltema2015} require either an  abundance of potassium far in excess of Solar, or an emissivity higher than  that predicted for these systems. 

{\bf The non-detection of such an X-ray line from dwarf spheroidal galaxies raises further questions about the nature of this emission line.  Dwarf spheroidal galaxies are extremely dark matter dominated objects and, given the observations of Bulbul et al.\cite{bulbul2014} and Boyarsky et al.\cite{boyarsky2014}, one expects to see X-ray emission from these objects as well.  However, recent simulations \cite{harada2014} show that it may be possible to account for the non-detection of X-ray emission from dwarf spheroidal galaxies using a multicomponent dark matter model with both cold and warm components, where the warm component is a sterile neutrino that decays via 3.5 keV photon emission and the cold component can be any heavy other sterile neutrinos, WIMPs or another dark-matter candidate.  There remains some controversy over the formation of such objects, since $\Lambda$CDM models over predict the formation of Milky Way satellites compared to current observations \cite{kravtsov2010,moore1999,Mathews14} whereas WDM models wash out dark matter over-densities on these scales \cite{schneider2014}.    These multicomponent dark matter simulations accurately describe the observed number of Milky Way satellites and the non-detection of X-ray emission from dwarf spheroidals. 

Moreover, we note that the constraint on the mixing angle in Figure \ref{fig:bounds} is based upon the assumption that a single sterile neutrino comprises 100\% of the dark matter.  If a decaying  sterile neutrino comprises only a small fraction of the total dark matter density due to other sterile neutrinos and/or WIMPS, etc., then a larger mixing angle is allowed.}

\section{Matter-enhanced neutrino oscillations \label{sec:osc}}

To solve the full neutrino mixing problem with a sterile neutrino requires an understanding of the full $4\times4$ mixing matrix, including mass differences and mixing angles between all neutrino species.  For the most part, however, studies have been limited to oscillations between just an electron neutrino $\nu_{e}$ and sterile neutrino $\nu_{s}$ (and their antiparticles) or just a $\nu_{\mu\tau}$ and sterile neutrino $\nu_{s}$.  

In vacuum, neutrino mixings between an electron neutrino and a sterile neutrino are governed by the vacuum mixing angle $\sin^{2} 2 \theta_{s}$ and mass-squared difference $\Delta m_{s}^{2}$.  However, in dense matter such as in supernovae, the forward scattering  of sterile neutrinos off of baryons and leptons gives rise to an effective ``in-medium'' mixing angle as a function of position $r$,
\begin{equation}
\sin^{2} 2 \theta_{M} (r) = \frac{\Delta^{2} \sin^{2} 2 \theta_{s}}{(\Delta \cos 2\theta_{s} - V_{\nu_{x}}(r))^{2} + \Delta^2 \sin^{2} 2\theta_{s}},
\label{eq:mixing}
\end{equation}
where $\Delta = \Delta m_{s}^{2}/(2 E_{\nu})$, $E_{\nu}$ is the neutrino energy, and $V_{\nu_{x}}(r)$ is the forward scattering potential for a neutrino of flavor $x$.  An electron neutrino in the supernova environment will experience a forward scattering potential of the form \cite{warren2014,caldwell2000}
\begin{equation}
V_{\nu_{e}}(r) = \frac{3 \sqrt{2}}{2} G_{F}\, n_{B} \left( Y_{e} + \frac{4}{3} Y_{\nu_{e}}+\frac{2}{3} Y_{\nu_{\mu}} + \frac{2}{3} Y_{\nu_{\tau}} - \frac{1}{3}\right),
\label{eq:potential}
\end{equation}
where $G_{F}$ is the Fermi coupling  constant, $n_{B}$ is the baryon number density, and $Y_{i}$ is the number fraction of species $i$.   A $\mu$ neutrino will experience  a  forward scattering potential of the form\cite{caldwell2000}
\begin{equation}
V_{\nu_{\mu}}(r) = \frac{ \sqrt{2}}{2} G_{F} n_{B} \left(Y_{e} +2 Y_{\nu_{e}}+4 Y_{\nu_{\mu}} + 2 Y_{\nu_{\tau}}-1\right)
\end{equation}
and a $\tau$ neutirino will experience
\begin{equation}
V_{\nu_{\tau}}(r) = \frac{ \sqrt{2}}{2} G_{F} n_{B} \left(Y_{e}  +2 Y_{\nu_{e}}+2 Y_{\nu_{\mu}} + 4 Y_{\nu_{\tau}}-1\right).
\end{equation}
The antineutrino species experience scattering potentials with the opposite sign.  In  the present work,  adopt  $Y_{\nu_{\mu}} = Y_{\nu_{\tau}} =0$, since $\mu$ and $\tau$ neutrinos and antineutrinos are produced entirely by pair creation processes and thus occur in equal numbers.

Looking at Eq.~(\ref{eq:mixing}), one can see that it is possible to get maximal mixing in matter ($\sin^{2} 2 \theta_{M} = 1$) even when the vacuum mixing angle is small.  This resonant mixing occurs for neutrinos with the energy
\begin{equation}
E_{res} = \frac{\Delta m^{2}}{2 V_{\nu_{x}}(r)} \cos{2 \theta}.
\label{eq:res}
\end{equation}
Thus, any neutrinos of species $x$ at position $r$ and with energy $E_{res}$ will experience a  Mikheyev-Smirnov-Wolfenstein (MSW) resonance \cite{mikheyev1985,wolfenstein1978}.

In this study, we have only considered coherent and adiabatic neutrino oscillations.  Coherent oscillations occur when the resonance width is greater than the neutrino mean free path.  The resonance occurs over a finite distance (or time) on the neutrino's world line,
\begin{equation}
\Delta r_{res} = \left| \frac{d \ln V_{\nu_{x}}(r)}{dr}\right|^{-1} \tan 2 \theta_{s},
\end{equation}
which defines the resonance width.  This corresponds to the distance over which the in-medium mixing angle falls from a maximum to one half ($\sin^{2} 2 \theta_{M} = 1/2$).  
  The adiabaticity of the resonance is a measure of the efficiency of the conversion of neutrino flavor.  The resonance can be considered adiabatic if the the resonance width is much greater than the oscillation length,
\begin{equation}
\gamma = \frac{\Delta r_{res}}{\ell_{osc}^{res}} \gg 1,
\end{equation}
where the oscillation length is related to the neutrino energy and vacuum mixing parameters, 
\begin{equation}
\ell_{osc}^{res} = \frac{4 \pi E_{res}}{\Delta m_{s}^{2} \sin 2 \theta_{s}}.
\end{equation}
If the oscillation is both coherent and adiabatic, all of the neutrinos of a given flavor $x$  within an energy band $\Delta E_{res} = E_{res} \tan 2 \theta_{s}$ centered at  the resonance energy $E_{res}$ will oscillate to sterile neutrinos, and vice versa.  

It is likely that decoherent, scattering-induced oscillations will be important in the high matter densities achieved in the collapsing core, but this will not be a dominant effect and it is sufficient here to only consider coherent and adiabatic neutrino oscillations.  We leave the exploration of scattering-induced effects to future work.

In Ref.~[\refcite{warren2014}]  the UND/LLNL \cite{book} spherically-symmetric general relativistic hydrodynamic supernova model with a 20~M$_{\odot}$ progenitor model from Ref.~[\refcite{woosley1995}]  was used to study the impact of sterile neutrino oscillations on the explosion energy.   A model that can explode successfully was used as a starting place and baseline for comparison.  For further details of the UND/LLNL supernova model, we refer the reader to Refs.~[\refcite{book,warren2014}].

\section{Simulations \label{sec:results}}

\begin{figure}[t!]
   \centerline{   \includegraphics[width=\textwidth]{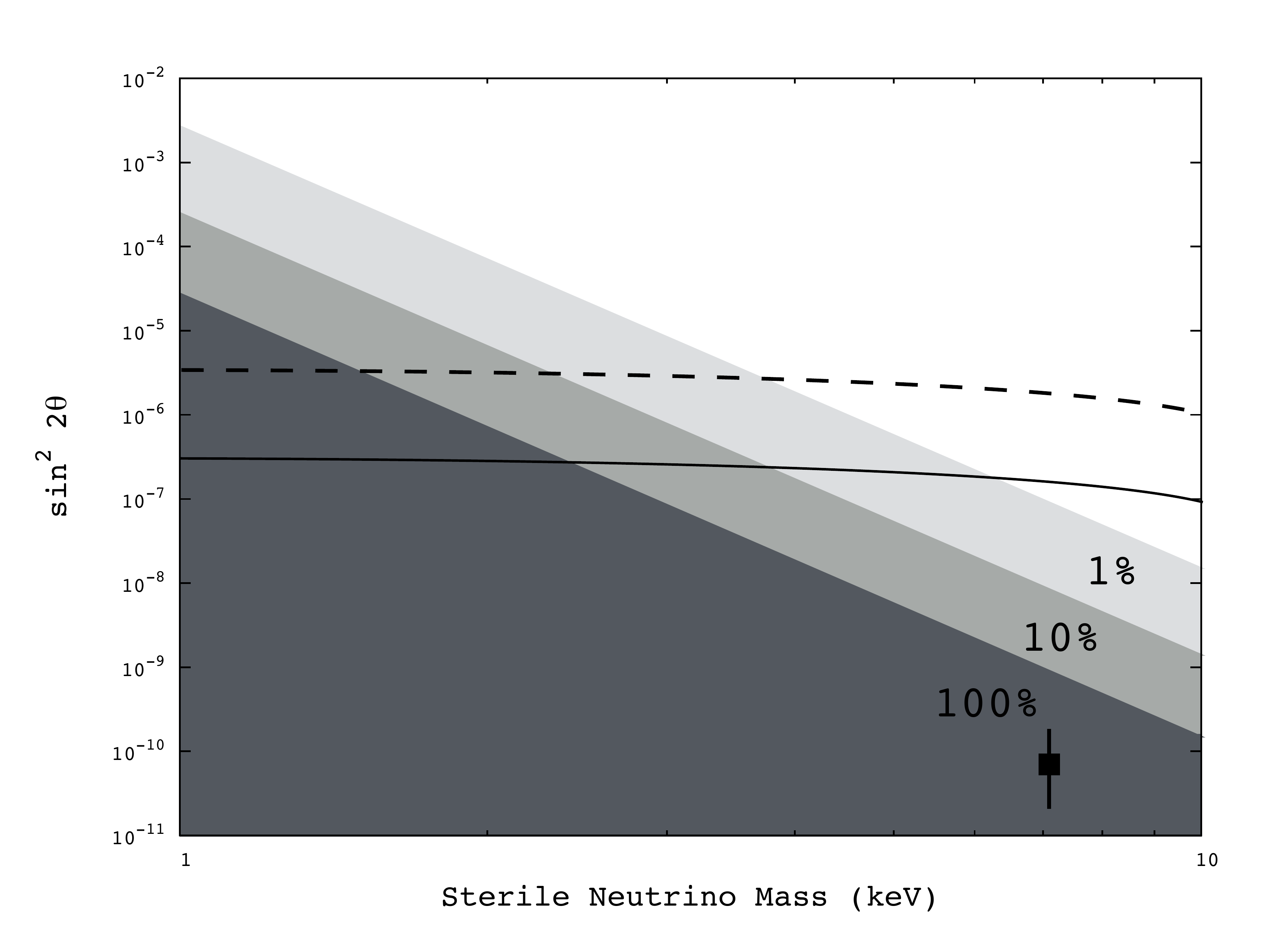} }
   \vspace*{8pt}
   \caption{Enhanced explosion energies for the sterile neutrino mass $\Delta m_{s}$ and mixing angle $\sin^{2} 2 \theta$ parameter space for $\nu_{e}\leftrightarrow \nu_{s}$ mixing.  The region above the solid line enhances the explosion energy by 1.01$\times$ compared to a simulation without a sterile neutrino, and above the dashed line is the region that enhances the explosion by $1.10\times$.  The shaded regions indicates the parameter space allowed for sterile neutrino dark matter if sterile neutrinos contribute 100\%, 10\%, and 1\% of the dark matter mass  \cite{abazajian2001a,boyarsky2006a,boyarsky2006b,boyarsky2007,boyarsky2009c} and the solid black square shows the most recent best fit point from the X-ray flux.\cite{boyarsky2014,bulbul2014}}
   \label{fig:parameter}
\end{figure}

 Figure~\ref{fig:parameter} shows results of simulations for a wide range of masses (1 keV $< \Delta m_{s} <$ 10 keV) and mixing angles ($10^{-11}< \sin^{2} 2 \theta < 10^{-2}$) which encompasses the region of parameter space that corresponds with dark matter candidates.  

For $\nu_{e}\leftrightarrow \nu_{s}$ mixing, Figure~\ref{fig:parameter} shows the enhancement to the explosion energy in a simulation with a sterile neutrino compared to the explosion energy in a simulation without a sterile neutrino.  The shaded region indicates the allowed parameter space for dark matter candidates.  The darkest region are the bounds if sterile neutrinos contribute 100\%  to the observed dark matter mass, the intermediate shading is for a 10\% contribution, and the lightest region shows the bounds for a 1\% contribution.  There is a large region of the parameter space that enhances the explosion energy by at least a few percent and part of this parameter space overlaps with  possible dark matter candidates.  However, the 7 keV sterile neutrino inferred from X-ray observations would not affect the supernova unless it has a substantial mixing angle ($\sin^2{(2 \theta)} > 10^{-7}$) corresponding to $< 1$\% of the total dark matter density.

 Similarly, for $\nu_{\mu\tau} \leftrightarrow \nu_{s}$ mixing, we found no significant enhancement (or depletion) to the shock energy for the parameter space of sterile neutrino mass and mixing angle considered here.  This is likely due to the fact that the $\mu-\tau$ neutrinos do not experience efficient matter-induced oscillation.

 Although we have shown that the scenario of a single sterile neutrino constituting 100\% of the dark matter does not affect the dynamics of CCSNe, one would like to know the 
possible effect of a lighter stable sterile neutrino or a 7 keV sterile neutrino with a much larger mixing angle contributing to a small fraction of the total dark matter.   {\bf For illustration, we also show how a lighter sterile neutrino might affect the supernova dynamics.  In particular, we show the effects of both a 1~keV and 7~keV sterile neutrino with a  mixing angle of $\sin^{2} 2 \theta_{s} = 10^{-5}$ for $\nu_{e}\leftrightarrow \nu_{s}$ mixing is adopted.  Both masses and this mixing angle maximize and  highlight the physics by which the shock reheating occurs.   We have observed in our simulations that any choice of mass and mixing angle that causes an enhancement in the explosion affects the supernova dynamics in roughly the same way except for the possible presence of episodic bursts described below.   Although both of these choices of mass and mixing angle are excluded by DM constraints, as per  Fig.~\ref{fig:bounds}, if a sterile neutrino contributes to 100\% of the DM density, in the more likely scenario that there are significant contributions to the total DM density from other sterile neutrino flavors, WIMPs, etc., these parameter values are allowed. For smaller mixing angles allowed by DM constraints, the mechanism will be the same, but much less dramatic.}

\subsection{$\nu_{e}\leftrightarrow \nu_{s}$ Mixing}

\begin{figure}[b!]
   \centerline{   \includegraphics[width=\textwidth]{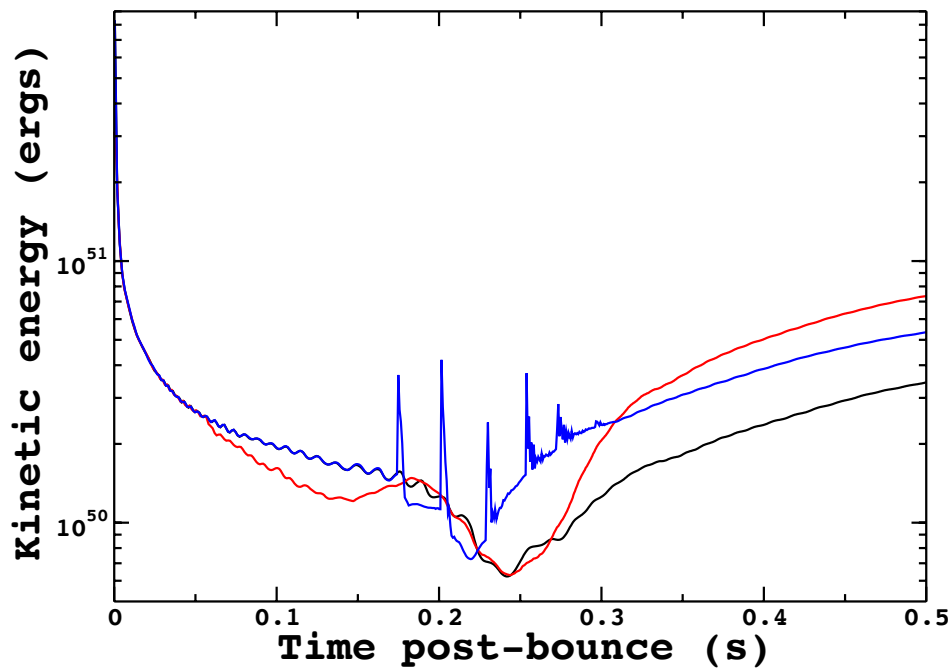} }
   \vspace*{8pt}
   \caption{Kinetic energy versus time post-bounce.  The black line is for a simulation without a sterile neutrino, the red line is for a sterile neutrino with mass $m_{s} = 1$~keV and mixing angle $\sin^{2} 2 \theta_{s} = 10^{-5}$, and the blue line is for a sterile neutrino with mass $m_{s} = 7$~keV and mixing angle $\sin^{2} 2 \theta_{s} = 10^{-5}$. (Color available online)}
   \label{fig:ke}
\end{figure}

In the case of $\nu_{e}\leftrightarrow \nu_{s}$ mixing, the enhancement to the explosion energy can be  quite dramatic if the sterile neutrino mixing angle is as large as $\sin^{2} 2 \theta_{s} = 10^{-5}$. This is evident in both the parameter plot (Fig.~\ref{fig:parameter}) and in more detail in Figure \ref{fig:ke}.  Figure \ref{fig:ke} shows the kinetic energy versus time post-bounce for simulations without a sterile neutrino and with either a 1~keV or 7~keV sterile neutrino.  At 1~s post-bounce, the explosion energy is enhanced by a factor of $\sim2$ with a 1~keV sterile neutrino and by a factor of $\sim1.5$ with a 7 keV sterile neutrino. Although the enhancement to the kinetic energy isn't evident until about 0.2~s  post-bounce, this is inherent to the delayed neutrino reheating mechanism.  After the shock stalls, it takes a fraction of a second for the neutrino reheating to become effective.

\begin{figure}[t!]
        \centerline{     \includegraphics[height=0.28\textheight]{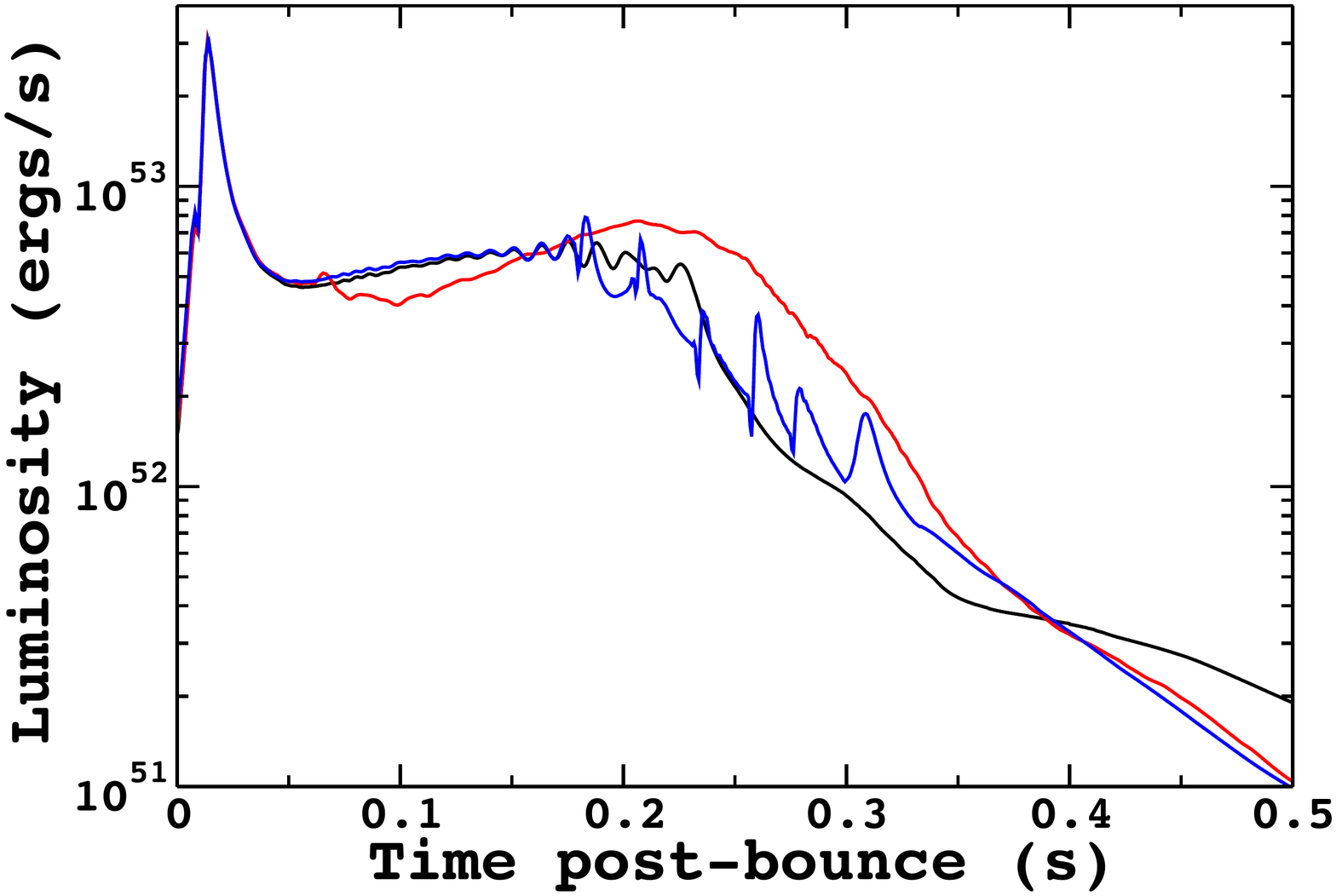} }
        \centerline{  \includegraphics[height=0.28\textheight]{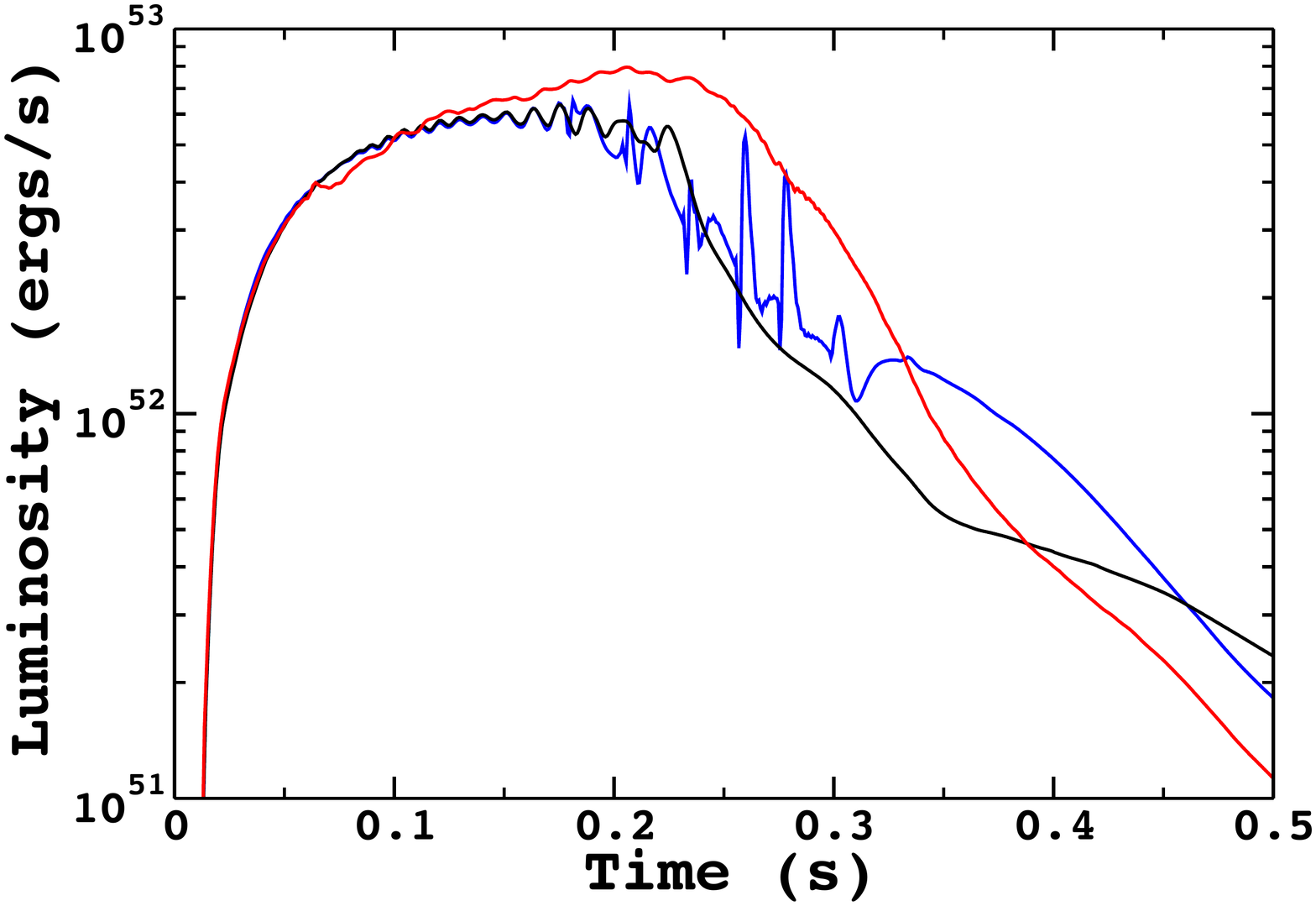}}
         \centerline{\includegraphics[height=0.28\textheight]{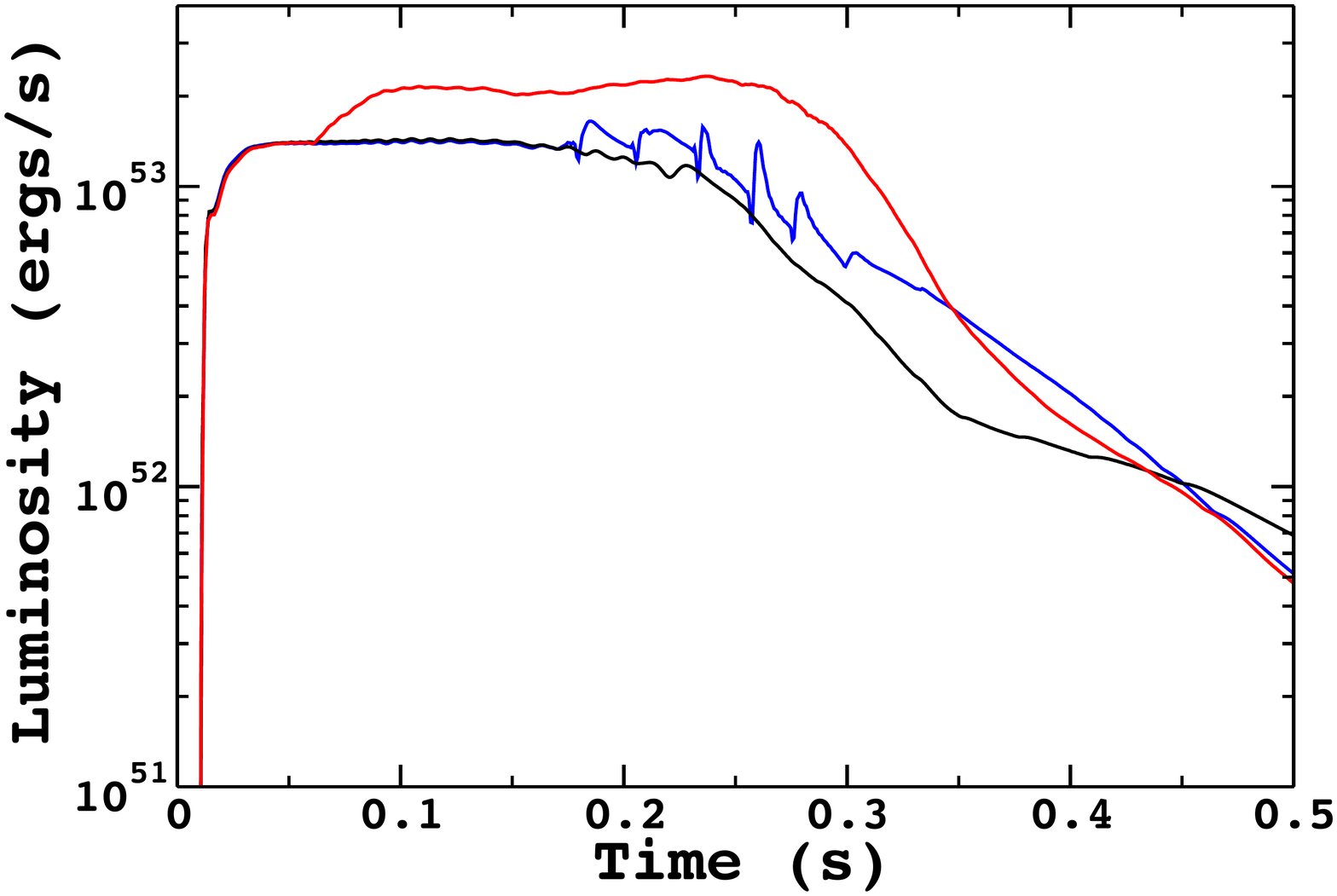}}
         \vspace*{8pt}
         \caption{Neutrino luminosity versus time post-bounce.  The black line is for a simulation without a sterile neutrino, the red line is for a sterile neutrino with mass $m_{s} = 1$~keV and mixing angle $\sin^{2} 2 \theta_{s} = 10^{-5}$, and the blue line is for a sterile neutrino with mass $m_{s} = 7$~keV and mixing angle $\sin^{2} 2 \theta_{s} = 10^{-5}$. The uppermost panel shows the electron neutrino $\nu_{e}$ luminosity, the middle panel shows the electron antineutrino $\bar{\nu}_{e}$ luminosity, and the lowest panel shows the $\mu-\tau$ neutrino $\nu_{\mu\tau}$ luminosity. (Color available online)}
        \label{fig:luminosity}
\end{figure}

The enhancement of shock reheating is due to an increase in the neutrino luminosities of all three neutrino types ($\nu_{e}$, $\bar{\nu}_{e}$, and $\nu_{\mu\tau}$).  Figure \ref{fig:luminosity} shows the neutrino luminosities versus time post-bounce, both with and without a sterile neutrino present.  The neutrino luminosities of all neutrino and antineutrino  flavors are enhanced from about 0.2s to 0.4s post-bounce.  It is this increase in the neutrino luminosities that increases the neutrino reheating behind the stalled shock, and thus enhances the explosion energy.  The timescale of the increase of the neutrino luminosities corresponds to the time post-bounce when the explosion energy becomes enhanced.

\begin{figure}[t!]
   \centerline{   \includegraphics[width = \textwidth]{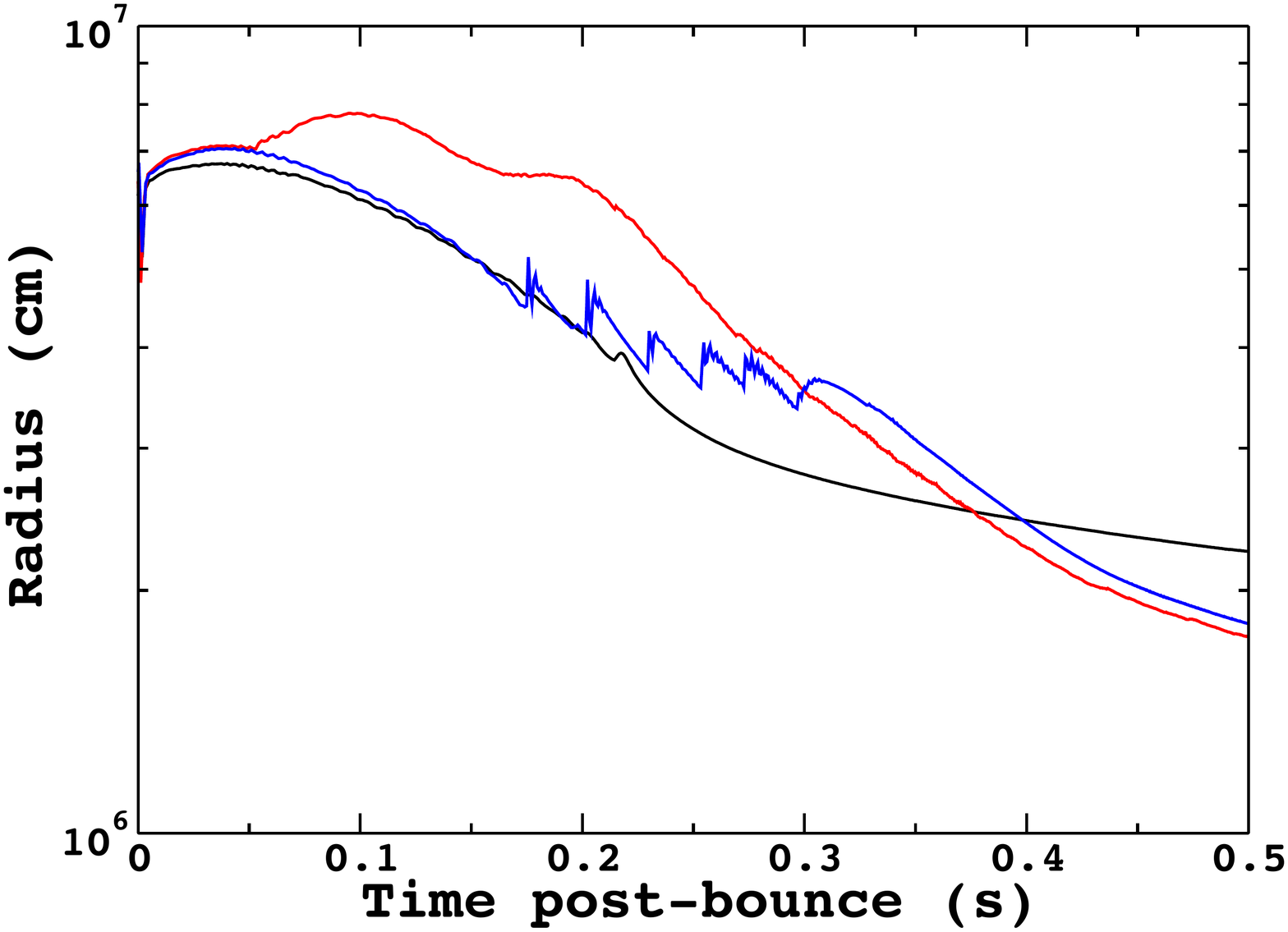} }
   \vspace*{8pt}
   \caption{Radius of the neutrinosphere versus time post-bounce.  The black line is for a simulation without a sterile neutrino, the red line is for a sterile neutrino with mass $m_{s} = 1$~keV and mixing angle $\sin^{2} 2 \theta_{s} = 10^{-5}$, and the blue line is for a sterile neutrino with mass $m_{s} = 7$~keV and mixing angle $\sin^{2} 2 \theta_{s} = 10^{-5}$.  (Color available online)}
   \label{fig:nusph}
\end{figure}

It is somewhat counterintuitive that oscillations between electron and sterile neutrinos would result in an increase in the luminosities of all neutrino flavors. However, it is important to note that the oscillation doesn't directly lead to increased neutrino reheating behind the shock.  Rather, a ``double reheating'' phenomenon occurs, whereby the $\nu_{e}\leftrightarrow \nu_{s}$ oscillation leads to additional heating near the location of the neutrinosphere, this heating at the neutrinosphere causes increased neutrino luminosities in all three flavors, and finally these enhanced neutrino luminosities reheat the stalled shock. 

\begin{figure}[t!]
   \centerline{\includegraphics[width=\textwidth]{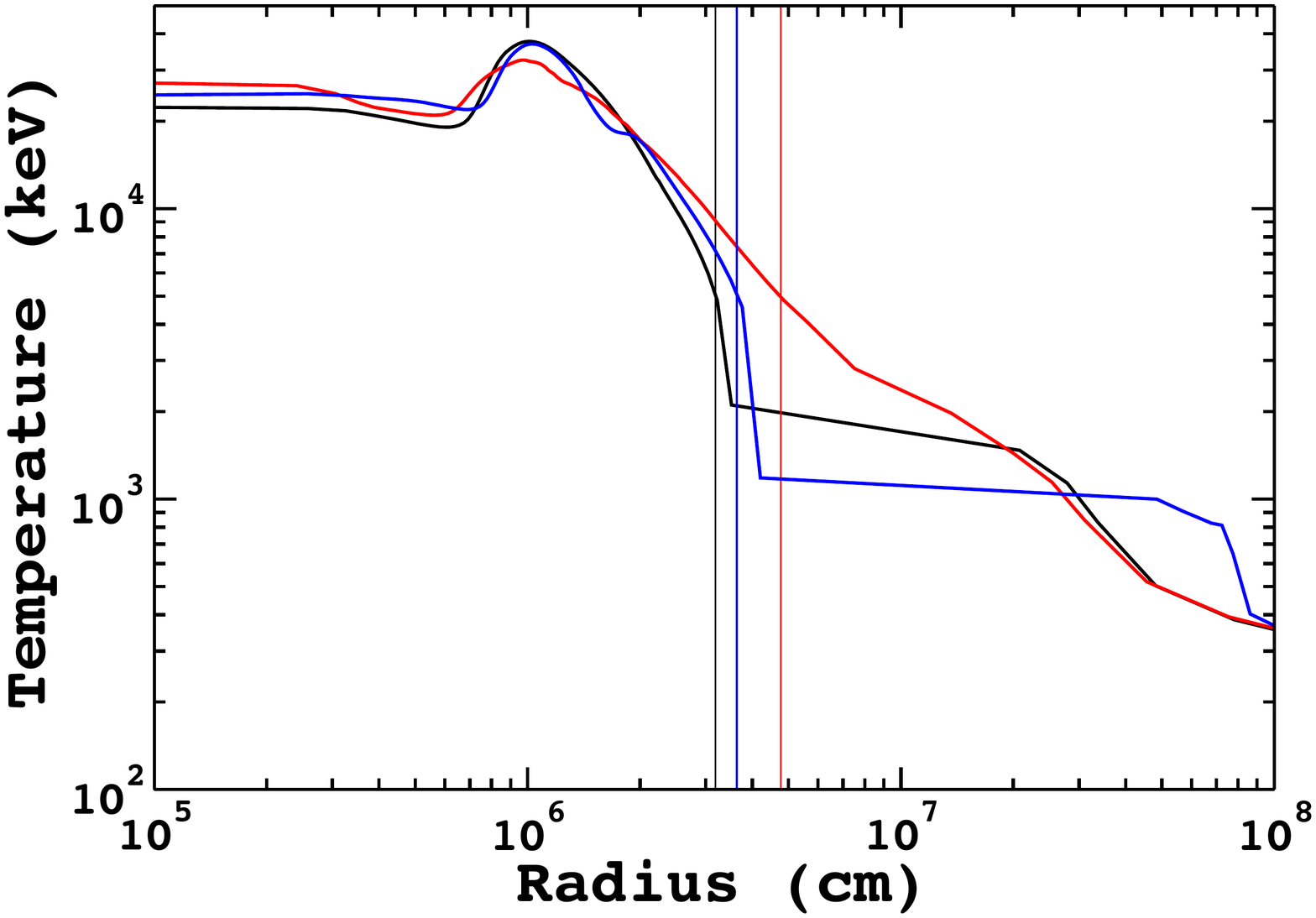} }
   \vspace*{8pt}
   \caption{Temperature versus radius at 0.25s post-bounce.  The black line is for a simulation without a sterile neutrino, the red line is for a sterile neutrino with mass $m_{s} = 1$~keV and mixing angle $\sin^{2} 2 \theta_{s} = 10^{-5}$, and the blue line is for a sterile neutrino with mass $m_{s} = 7$~keV and mixing angle $\sin^{2} 2 \theta_{s} = 10^{-5}$. The vertical lines represent the location of the neutrinosphere for the simulation with a 1~keV sterile neutrino (red line), simulation with a 7~keV sterile neutrino (blue line), and simulation without a sterile neutrino (black line). (Color available online)}
   \label{fig:nutemp}
\end{figure}

The enhancement of the neutrino luminosities in Figure \ref{fig:luminosity} based upon a 1~keV sterile neutrino are similar to those of Ref.~[\refcite{warren2014}] made with an $\approx 5$ keV sterile neutrino with similar mixing angle  with one exception.   A striking feature in the 5 keV case of Ref.~[\refcite{warren2014}] and the 7~keV case explored in this work is the appearance of episodic neutrino bursts  with a period of 30-40 ms (5 keV) or 40-50 ms (7 keV) in the luminosity and flux.  
This episodic variation of the neutrino luminosity is due to the fact that the neutrino photospheric luminosity of the proto-neutron star  is fixed  by the ratio of the total internal energy to the neutrino  diffusion time.  When the neutrino chemical potential falls above the resonance energy, those neutrinos in the energy groups corresponding to the resonance energy and width immediately have a diffusion time scale drastically shortened by the free streaming of the sterile neutrinos to just below the neutrinosphere.  Hence, there is a precipitous drop in the average diffusion time scale and a concomitant increase in the luminosity.  

However, in the 5 keV and 7 keV cases, once a significant fraction of available (mostly anti-electron) neutrinos is depleted, the process is shut off and the luminosity actually decreases until neutrinos can diffuse back into the depleted energy and spatial groups.   The difference between the 5 keV and the 1 keV case is due to the fact  that the resonance width $\Delta E_{res}  \propto \Delta m_s^2$.  Hence, the energy width of the depleted neutrino energy spectrum is much less for a 1 keV sterile neutrino than for a heavy 5 or 7 keV neutrino.  The 1 keV  neutrinos can quickly diffuse to fill the gap in the spectrum and the episodic burst effect is minimal.  Moreover, the amplitude and period of the luminosity spikes scale roughly linearly with neutrino mass.  This means that should an oscillation of the type shown here and in Ref.~[\refcite{warren2014}] may be detected in a next generation neutrino detector such as Hyper-kamiokande,\cite{abe11} its amplitude and period could be used to infer the mass of the oscillating sterile neutrino. 

 Independently of  the episodic fluctuation effect, the  basic mechanism for the enhancement in the observable neutrino luminosities is evident in the behavior of the neutrinosphere.  Figure~\ref{fig:nusph} shows the neutrinosphere radius versus time post-bounce in simulations with and without  a sterile neutrino.  It is evident that oscillations between electron neutrinos and sterile neutrinos heat the  proto-neutron star surface and cause it to expand by depositing additional neutrino energy at the location of the second resonance.  This leads to an increased neutrinosphere radius.  Figure~\ref{fig:nutemp} shows the temperature versus radius at an illustrative time (250 ms post-bounce) for simulations with and without a sterile neutrino.  As can be seen in Figure~\ref{fig:nutemp}, the  proto-neutron star surface is heated and extended in the simulation with a sterile neutrino, compared to a  simulation without a sterile neutrino.  Although the radius of the neutrinosphere is increased in the presence of a sterile neutrino, the temperature at the neutrinosphere remains the same in all three simulations.  Thus, the larger neutrinosphere area ($4 \pi R_{\nu}^{2}$) leads to an enhanced emission of neutrinos, but the neutrinos are emitted with nearly the same average energy in both simulations.
 
This figure confirms that scillations between an electron neutrino and sterile neutrino can lead to a significant enhancement of the explosion energy in core-collapse supernova simulations.  This enhancement is due to the expansion of the neutrinosphere and increased luminosities of all three neutrino types ($\nu_{e}$, $\bar{\nu}_{e}$, and $\nu_{\mu\tau}$).

\section{Conclusions \label{sec:conc}}

X-ray observations of galaxies and galaxy clusters indicate a possible emission line at $\sim$ 3.5 keV.  There are two proposed explanations of this line: atomic emission from potassium or a decaying 7~keV sterile neutrino dark matter.  The potassium lines near 3.5~keV are difficult to fit in the observed environments due to the uncertainties in the plasma temperature, backgrounds, and atomic abundances.  Many fits require a potassium abundance well  in excess of the Solar values.  This seems unlikely in the observed environments.

The observed X-ray emission may be due to decaying sterile neutrino dark matter.  If this is the case, bounds can be placed on the sterile neutrino mass and mixing angle from the photon energy and observed flux.    Further observations are needed to confirm the presence of this line in additional dark matter dominated objects, such as dwarf spheroidal galaxies.  {\bf If this observation indeed corresponds to a 7 keV sterile neutrino, we find that the inferred mixing  angle for a single sterile neutrino making up 100\% of the dark matter would imply almost no observable effect on core collapse supernovae.  However, if the 7 keV neutrino represents only a small ($< 1$\%) of the total dark matter density, it will alter the supernova dynamics and observable luminosities.  We note that such a neutrino is not ruled out by the constraints of Figure \ref{fig:bounds}.}

Thus, it remains possible that sterile neutrinos may play a part in a multi-component dark matter model, but further work is required to fully understand the observational signatures of such multi-component models.  As far as the impact on the CCSN explosion is concerned, a  multi-component model is required  to  accommodate   a sterile neutrino mass and mixing angle  that  leads to an enhanced explosion.  if the decaying sterile neutrino makes up ($< 1$\%) of  the dark matter density, a shorter decay lifetime is allowed corresponding to the larger mixing angle needed to impact the explosion.

Indeed, for oscillations between an electron neutrino and sterile neutrino, a large region of the sterile neutrino mass-mixing angle parameter space is allowed by these observations as long as the decaying sterile neutrino is a small fraction of the total DM density.  The larger associated mixing angle then leads to an enhancement in the explosion energy of CCSNe.  This enhancement to the explosion energy is mainly due to the heating and expansion of the neutrinosphere, which results in the increased luminosities of all neutrino (and antineutrino) flavors.  

We have also considered  $\nu_{\mu\tau} \leftrightarrow \nu_{s}$ oscillations.  However, for this case we find that there is little effect on the explosion energy.  This is because the forward scattering potential does not sufficiently drop below the neutrino chemical potential to induce significant  mixing.

 Nevertheless, a small contribution to the total dark-matter density from  keV mass neutrinos with a large mixing angle ($\sin^2{(2 \theta)} > 10^{-5}$) is  especially promising when one considers the current 
state of the art of 2- and 3-dimensional core-collapse supernova simulations.  Most multidimensional models explode via a hydrodynamic instability such as SASI or neutrino-heated convection, but the explosion energies tend to be lower than that required by observations.  Based on studies to date, if a sterile neutrino with a large mixing angle were included in a multidimensional model, its presence may be sufficient to enhance the explosion energy and thereby match observations.  Clearly, further work is merited to explore this possibility.

\appendix

\section*{Acknowledgments}

Work at the University of Notre Dame was supported by the U.S. Department of Energy under Nuclear Theory Grant DE-FG02-95-ER40934. 
One of the authors (MW) was also partially supported by the National Science Foundation under Grant No. PHY-1430152 (JINA Center for the Evolution of the Elements).
This work was also supported in part by Grants-in-Aid for Scientific Research of the JSPS (20105004, 24340060).

\end{document}